\documentclass[%
  12pt,
  paper=letter,
  abstracton,
  pagesize=auto,
  version=last,
  DIV=calc
  ]{scrartcl}
\typearea[current]{calc}
\areaset[current]{\textwidth}{\textheight}
\usepackage[english]{babel}
\usepackage{datetime}
\renewcommand{\dateseparator}{-}
\newcommand{\todayiso}{\the\year \dateseparator \shortmonthname \dateseparator \twodigit\day}
\usepackage[letterpaper,margin=1in]{geometry}
\usepackage[T1]{fontenc}
\usepackage{ae,aecompl,aeguill}
\usepackage{fix-cm}
\usepackage{lmodern}
\usepackage{microtype}
\usepackage{ccicons}    
\usepackage{ucs}
\usepackage[utf8x]{inputenc}
\usepackage[x11names]{xcolor}
\usepackage{graphicx}
\DeclareGraphicsExtensions{.pdf,.gif,.png,.jpg}
\usepackage{xspace}
\usepackage{setspace}
\usepackage[colorlinks=true,urlcolor=blue]{hyperref}
\hypersetup{
  pdftitle    = {Gauge Invariance and the Goldstone Theorem},
  pdfauthor   = {G.S. Guralnik <gerry@het.brown.edu>},
  pdfsubject  = {Symmetry Breaking, High Energy Theory, Feldafing,
    1964, GHK, Goldstone, Gauge, Invariance, Symmetry, Guralnik,
    Hagen, Kibble},
  pdfcreator  = {LaTeX2e with hyperref package},
  pdfproducer = {pdflatex},
  pdfkeywords = {Gauge, Quantum, Symmetry, Breaking, GHK, Feldafing,
    Goldstone, Invariance, GHK, 1964, Guralnik, Hagen, Kibble},
  pdfview     = {FitH}
}
\usepackage{cite}
\usepackage{xfrac,xkeyval}
\usepackage{amsmath,amsfonts,mathrsfs,dsfont,amstext,amssymb,amsbsy,amsopn,amsthm,eucal}
\usepackage{mathtools}
\DeclareMathAlphabet{\mathpzc}{T1}{pzc}{m}{it}   
\usepackage{mathrsfs}
\usepackage{empheq}
\usepackage{textcomp}
\usepackage{wasysym}
\usepackage{fancyhdr}
\fancypagestyle{plain}{%
  \fancyhf{}
  \fancyfoot[C]{\bfseries\thepage}

}
\newcommand{\helv}{\fontfamily{phv}\fontseries{b}\fontsize{9}{11}\selectfont}
\makeatletter
\def\equalsfill{$\m@th\mathord=\mkern-7mu
  \cleaders\hbox{$\!\mathord=\!$}\hfill
  \mkern-7mu\mathord=$}
\makeatother

\def\DHLhksqrt#1#2{\setbox0=\hbox{$#1\sqrt{#2\,}$}\dimen0=\ht0
  \advance\dimen0-0.2\ht0
  \setbox2=\hbox{\vrule height\ht0 depth -\dimen0}%
{\box0\lower0.4pt\box2}}
\def\blfootnote{\xdef\@thefnmark{}\@footnotetext}
\long\def\symbolfootnote[#1]#2{\begingroup%
  \def\thefootnote{\fnsymbol{footnote}}\footnote[#1]{#2}\endgroup}
\begin{document}
%
\begin{flushright}
  {\textsf{BROWN-HET-\textbf{1610}}}
\end{flushright}
\bigskip

\begin{center}
  {\Large \textbf{\textsf{Gauge Invariance and the Goldstone Theorem}}} \\
  \vspace{0.75in}

  {\Large \textsf{Gerald S.~Guralnik\footnote{\href{mailto:gerry@het.brown.edu}{\texttt{gerry@het.brown.edu}}}}} \\
  \bigskip\bigskip

  \textsf{Physics Department\footnote{\href{http://www.het.brown.edu/}{\texttt{http://www.het.brown.edu/}}}.}\\

  \textsf{Brown University, Providence --- RI. 02912}
\end{center}
\bigskip\bigskip\bigskip

\begin{center}
  \textbf{\textsf{Abstract}}
\end{center}
\bigskip
\noindent This manuscript was originally created for and printed in the
``Proceedings of seminar on unified theories of elementary particles''
held in Feldafing Germany from July 5 to 16 1965 under the auspices
of the Max-Planck-Institute for Physics and Astrophysics in
Munich. It details and expands upon the Guralnik, Hagen, and Kibble
paper that shows that the Goldstone theorem does not require
physical zero mass particles in gauge theories and provides an example through the model which has
become the template for the unified electroweak theory and a main component of the Standard Model.

\newpage
\begin{center}
  \textbf{\textsf{Introduction}}
\end{center}
\bigskip

This manuscript was created for and printed in the ``Proceedings of
seminar on unified theories of elementary particles'' held in Feldafing
Germany from July 5 to July 16 1965 under the auspices of the
Max-Planck-Institute for Physics and Astrophysics in Munich. It is
based on the talk that I gave at this conference which was essentially
identical to earlier talks at Imperial College (then my home
institution) and Edinburgh. It expands and details the arguments given
in the Guralnik, Hagen and Kibble paper \cite{13} demonstrating that
the Goldstone Theorem does not require a physical massless particle in
gauge theories  and examines the explicit scalar electrodynamic model
which now forms the basis of the unified model of weak and
electromagnetic interactions and inspired ``Higgs'' searches at LEP, the
Tevatron and the LHC. This paper shows how the arguments evolved from
my earlier PRL paper \cite{10}. As in the Guralnik, Hagen and Kibble
paper, this paper keeps close track of the physical degrees of freedom
and explicitly shows, in leading order of the model, that a massive
vector boson and a scalar boson (whose mass is generated by higher
order contributions) describe the entire excitation spectrum. In the
radiation gauge solution of this model the  Goldstone theorem is not
valid and consequently imposes no constraints requiring zero mass. The
Lorentz gauge solution, which yields the same physical results, obeys
the Goldstone theorem (as it must) by having irrelevant massless gauge
excitations. The original work was supported by the U.S. National
Science Foundation through a postdoctoral fellowship. Except for this
introduction, the acknowledgements and the associated header information, I have attempted to
make this article identical in content to the original complete with
some potentially confusing notation, typographic errors and (minor) errors in language and physics. I am posting this here because I have been receiving requests for copies since the original proceedings are
not readily available.
\bigskip\bigskip

\vspace{0.5in}
\begin{center}
  \textbf{\textsf{Original Feldafing Conference Paper}}
\end{center}

The impressive success of the method of broken symmetries in
non-relativistic problems have understandably led to the hope that
similar techniques might be the key to at least some of the problems
of the relativistic theory.  The early calculations \cite{1}~\cite {2}
demonstrated that the requirement that the vacuum expectation value of
a field operator be non-vanishing, could indeed lead to solutions of
field theoretic problems not realized by normal perturbative methods
while not drastically destroying the normal structure expected of a
relativistic field theory.  Nevertheless, to date, these methods have
not resulted in any particularly gratifying insights into nature
except for a possible indication of how the photon can be regarded as
a composite particle \cite{3}.  The desirable guarantee that the
photon has zero mass which naturally comes from this technique, in
fact, seems to be reflective of a limitation which prohibits its
application to a wide range of problems.  Indeed, it was realized at
the initial stages that a broken symmetry might always be associated
with a zero mass particle.  General proofs of this were given by
Goldstone, Salam and Weinberg, and Bludman and Klein~\cite{4}.

To emphasize our point of view we give here a proof which is
completely equivalent to the usual proofs, but makes points essential to
our discussion more transparent.  Assume that we are given some time
independent generator $Q$ on fields $\varphi_i$ such that

\begin{equation}
\label{eq:1}
\left[Q, \varphi_{i}(y)\right] = t_{ij}\, \varphi_{j}(y)
\end{equation}
and

\begin{equation}
  \label{eq:2}
  Q = \int d^{3}x j^{0}(x) \; .
\end{equation}
To ``break'' the symmetry $Q$ the requirement

\begin{equation*}
t_{ij} \, \langle 0|\varphi_{j}|0\rangle = n_{i} \neq 0
\end{equation*}
is imposed. That $n_i$ is independent of $y$ combined with
energy momentum conservation gives a definite statement of the nature
of the excitation spectrum of $\varphi_i$. In this case

\begin{equation*}
  \langle 0| \left[j^{0}(x), \varphi_{i}(y)\right] |0\rangle =
    n_{i}\int \frac{e^{ip(x-y)}}{(2\pi)^3}\, G(\vec{p}, p^{0}) d^{4}p,
\end{equation*}
so the application of \eqref{eq:1} and \eqref{eq:2} yields

\begin{equation}
  \label{eq:3}
  \int d^{3}p\, \delta^{3}(\vec{p})\, G(\vec{p}, p^{0}) = \delta(p^{0})
\end{equation}

If it is assumed that the theory is relativistically invariant and
that the broken symmetry requirement in no way interferes with the
relativistic structure, equation \eqref{eq:3} requires that $G(\vec{p},
p^{0}) = \delta(p^{2})\, p^{0}\, \epsilon(p^{0})$ + possible
irrelevant terms.  Thus we conclude that $\varphi_i$ excites a zero
mass particle.  This remarkably simple result is the Goldstone
theorem.  It is the only known exact statement on the excitation
spectrum of a field operator and consequently has been viewed with
suspicion since its inception.  There is, however, little question
that these results are correct when the large number of implicit
existence assumptions which have gone into the proof and which are
usually valid in other contexts hold here as well.  The fact that the
Goldstone bosons of one useful broken symmetry theory, the
superconducting electron gas, are massive, has motivated considerable
interest in how to get around the assumptions of the theorem without
too drastically mutilating the underlying field theoretic structure
which made its proof possible \cite{5}.

Historically, the first attempts were made with the belief that even
in non-relativistic theories the assumptions of the theorem are still
valid but that "spurious states" enter to avoid the zero mass particle
conclusion.  The basic observation \cite{6} is that non-relativistic
theories may be imbedded in a relativistic theory by the introduction
of a time-like vector $\lambda^{u}=(1,0,0,0)$.  A straighforward
analysis demonstrates that Fourier transform

\[
  \langle 0 |\left[j_{u}
  (x)~\varphi(y)\right]| 0\rangle = (A + B\epsilon(k^{0}))k_{u}\,
  \delta(k^{2}) + C\, \lambda_{u}
  \delta^{3}(k)\, \delta(\lambda\cdot k) + [k_{u}(\lambda\cdot
  k) - \lambda_{u}k^{2}]\, D (k^{2}, \lambda\cdot
  k) + E\, \lambda_{u}\, \delta(\lambda\cdot k) \; .
\]

The term proportional to $C$ is the one on which attention was
initially focussed. It is most unusual in that it represents an
isolated state that might be interpreted as a transition between the
various degenerate vacua which are formed as a consequence of breaking
the symmetry.  In fact, though this term can simulate the behavior of
the global generator $Q$, it cannot consistently be interpreted as
contributing to the density commutator since
$\langle 0 |[{j^{\circ}}(x)\varphi(y)]|0\rangle\big|_{{x^{0}=y^{0}}}
\propto \delta^3(\vec{x}-\vec{y}) +$ possible irrelevant terms
while

\[
  \int d^4k \,e^{ik(x-y)}\, C\, \lambda^{0}\, \delta^3(k)
  \delta(\lambda\cdot k)\big|_{x^{0}=y^{0}} = C\, \big|_{k^{0}=0}\; .
\]

Consequently, no escape from the Goldstone theorem can ever occur in
this manner.  This point has been made in rather different language by
Lange \cite{7}. The ``spurious'' contribution proportional to $E$ cannot
be disposed of, nor is its interpretation quite so simple.  It is the
sort of term that one is not likely to obtain when using usual
approximation methods, and to my knowledge no model exists where this
term appears when the calculation is properly handled.  In principle,
however, there is no reason to believe that a simple non-relativistic
model demonstrating this transition cannot be formulated.  Also, such
transitions might be important in apparently relativistic theories
where Lorentz invariance is broken.  The Bjorken theory in its
original current non-conserving approximation is an admittedly poor
example of where such terms appear in the check of the Goldstone
commutator \cite{8}.

The third term is the one that is important for the understanding of
how the Goldstone theorem can break down.  Indeed such a term appears
in broken symmetry theories involving vector gauge fields \cite{9}.
As written it seems innocent enough, but it must be appreciated that
the particle associated with such a term will always be massive and
will necessarily involve the breakdown of the global conservation law
and hence a negation of the usual assumptions made to prove the
Goldstone theorem.  Thus, this term will give results entirely
different in nature from those derived from massless particles of
``spurions''.  We shall return to this important point later.

It is our object first to demonstrate how the old arguments on the
connection between gauge invariance and zero mass can be expressed in
an analytic way through the use of the Goldstone theorem \cite{10}.
Then by studying the breakdown of this connection, we will have a guide
toward the understanding of how the method of broken symmetries need
not lead to massless physical particles.

Consider the simplest possible field theory, that of a massless free
spinless particle described by the Lagrangian

\[
  L=\varphi^u \partial_u \varphi + \frac{1}{2} \varphi^u \varphi_u \; .
\]

Note that since there is no mass term, this $L$ possesses the gauge
invariance $\varphi \rightarrow \varphi+n, \varphi^u \rightarrow
\varphi^u$. It is our intention to display an operator realization of
this transformation.  This is easily done since from the field
equation $\partial_u \varphi^u=0$, it follows that

\begin{equation*}
L_n\equiv \int d^3x \,n \,\varphi^{0}(x)
\end{equation*}
with $n$ constant is independent of time.  In fact $L_n$ does not
exist unless $n$ is adjusted to fall off rapidly to zero for large
spatial coordinates.  We shall, however, proceed formally with $n$
constant, as ignoring this subtlety does not get us into any
difficulties.  Those who object to this procedure may find comfort
either by noting that all our results are derived by considering well
defined commutators of $L_n$ or better still be referring to the paper
of Streater which treats these problems rigorously \cite{11}.  Since
the canonical commutation relations have the form

\begin{equation*}
i\, [\varphi^{0} (x), \varphi(y)]_{x^{0}=y^{0}} =  \delta^3(\vec{x} - \vec{y}),
\end{equation*}
it follows that

\begin{equation*}
i\, [L_n, \varphi(y)] = n \; .
\end{equation*}
In particular, introducing the usual set of states$|a'\rangle$ for which
$\langle 0|\varphi|0\rangle = n$, it follows that

\begin{equation*}
i\, \langle 0|[L_n, \varphi(y)]|0\rangle = n \; ,
\end{equation*}
and consequently the Goldstone theorem states that $\varphi(y)$
excites a massless particle.  This of course, is a completely trivial
and circular observation for this model.  Now these conclusions may be
applied to make this example look like a broken symmetry theory.  To
do this, we introduce a new set of states defined by the relation

\begin{equation*}
|n a^{\prime}\rangle \equiv e^{-i L_n} |a^{\prime}\rangle.
\end{equation*}

Although, formally it appears that the states $|n\, a^{\prime}\rangle$ are
unitarily connected with the states $|a^{\prime}\rangle$ this is not the
case, as $L_n$ is properly defined by a limiting process where
$n(\vec{x})$ is taken to be constant over an increasingly large
spatial volume.  As this volume becomes infinite, $|a^{\prime}\rangle$ and $|
n a^{\prime}\rangle$ become members of different inequivalent
representations.  In this limit all state vectors of one set become
orthogonal to all the state vectors in the other set.  It is
emphasized that no finite product of operators $\varphi(y)$ can induce
a transition between members of different sets of inequivalent
representations, and consequently in this special type of theory no
``spurions'' occur as an expression of such transitions.

It follows at once that
\begin{align*}
\langle b^{\prime} n |\varphi| n a^{\prime}\rangle &= \langle b^{\prime} |\varphi
  |a^{\prime}\rangle + i \langle b^{\prime}| [L_n, \varphi]|a^{\prime}\rangle \;.\\
&= \langle b^{\prime} |\varphi |a^{\prime}\rangle + n \langle
  b^{\prime}|a^{\prime}\rangle \; .
\end{align*}
In particular $\langle 0 n | \varphi | 0 n\rangle = n$, so a non-vanishing
expectation value of a field operator has been realized through this
technique explicitely because of the natural presence of zero mass
particles.  The realization has occurred through the possibility of
constructing states differing from the original states through the
addition of an infinite number of zero energy, zero mass particles to
these states.

To make full contact with the usual formulation of broken symmetry
theories, assume that the field $\varphi$ has two degrees of freedom

\begin{equation*}
\varphi = \left( \begin{array}{c} \varphi_1\\ \varphi_2 \end{array} \right),
\end{equation*}
then there is a conserved bilinear local current $j^u = i \varphi^u\,
q\, \varphi$ with the corresponding global charge $Q = \int d^3 x j^0(x)$.  Since

\begin{equation*}
[Q, \varphi (x)] = q\, {\varphi (x)},
\end{equation*}
it follows that

\begin{equation*}
\langle 0 n | [Q, \varphi(x)] | n 0\rangle = q \,n \; ,
\end{equation*}
and consequently there is a Goldstone theorem in the usual sense
involving a bilinear generator.  In this simple case, as always, the
theorem is nothing more than an expression of the possibility of
making constant gauge transformation on a massless field.  However,
since here the canonical group is the exact carrier of the broken
symmetry this statement has a particularly nice realization through
the decomposition

\begin{equation*}
\langle 0 n | [Q, \varphi(x)] | n 0\rangle = i \int d^3 x \langle 0 | [ \varphi^{0}\, q\, n, \varphi] | 0\rangle
= i \langle 0 | [ L\,_q\,_n, \varphi ] | 0\rangle\;.
\end{equation*}

One point should again be emphasized.  The fact that $\langle 0 | [L\,_q\,_n,
\varphi] | 0\rangle$ is non-vanishing is an expression of a fundamental
degeneracy of any system with zero mass particles.  It is a
demonstration that in the system under consideration there are states
carrying unit charge relative to the vacuum but of the same energy.
On the other hand, though the non-vanishing of the commutator $\langle
0 n | [Q, \varphi (x)] | n 0\rangle$ is directly due to the realization of such a
degeneracy, the states $\langle a^{\prime} n|$ should not be looked at as being
any of these states.  It is clear from their definition that
$\langle a^{\prime} n |$ are not eigenstates of the operator $Q$, but are in
fact a superposition of states of all possible charges.  They are
orthogonal to the states $|a^{\prime}\rangle$, but form the basis for a
physically equivalent description in which the observable current is
measured relative to the vacuum and is $j^{u^{\prime}} = i[\varphi^u
q\, \varphi - \varphi^u\, q\, n]$.  It is consequently clear that this type
of model does not really ``break'' the symmetry.

The considerations here have been of the most elementary sort,
nevertheless they serve as a basis towards the understanding of the
Goldstone theorem.  The results become much less trivial if we add to
the free Lagrangian any interaction whatsoever depending on
$\varphi^u$ and any other fields except $\varphi$.  In this case all
the above statements are still valid.

Now we proceed to the only slightly more complicated problem of free
electrodynamics.  In the radiation gauge the field equations are

\begin{equation*}
F^{u v} = \partial^u A^v - \partial^v A^u
\end{equation*}
and

\begin{equation*}
\partial_v F^{u v} = 0
\end{equation*}
with the equal time commutation relations

\begin{equation*}
[F^{0 k} (x), A^l (y)]_{x^{0} = y^{0}} = i [ \delta_{k l} - \frac{\partial_k \partial_l}{\nabla^2}] \delta^3 (\vec x - \vec y).
\end{equation*}
From the field equations it follows that

\begin{equation*}
L_n = \int d^3x\, F^{0k}\, n_k
\end{equation*}
is independent of time.  The commutation relations establish that
$\frac{1}{i} [L_n, A^k]=\frac{2}{3} n^k$, and consequently just as in
the scalar model there is a Goldstone theorem and the possibility of
manufacturing a set of states $| n a^{\prime}\rangle \equiv e^{iL_n} |
a^{\prime}\rangle$ for which $\langle a^{\prime} n | A^k | n b^{\prime}\rangle =
\langle a^{\prime} | A^k | b^{\prime}\rangle + \frac{2}{3} n^k$.

In the Lorentz gauge the field equation is

\[\partial_u (\partial^u A^v) = 0\; ,\]
while usually the canonical commutation relation is taken as

\begin{equation}
\label{eq:4}
[\partial_0 A^v (x), A^u (y)]_{x^0=y^0}=-i g^{u v} \delta^3 (\vec x - \vec y) \; .
\end{equation}
From this time independent generator $L_n=\int d^3 x \partial_0 A^u
(x) n_u$ is formed which satisfies the relation
\[i [L_n, A^u (y)] = n^u \; .\]

The states $|na^{\prime}\rangle$ are formed in the usual manner.  The
bilinear symmetry which is broken here as in any theory in which the
field whose expectation is non-vanishing carries intrinsic spin is
Lorentz invariance.  In more complicated theories that breaking of
this symmetry is extremely dangerous, because of the possibility of
the realization of a non-covariant excitation spectrum.

Now consider interacting electrodynamics in the radiation gauge.  All
is the same as before except that the interaction is inserted by
replacing the free equation for $F^{uv}$ by the equation

\[\partial_u F^{uv}=-e_0 j^v\; .\]

Since this may be rewritten as $\partial_u[F^{uv} + e^0 x^v j^u]=0$,
application of the same reasoning as in the preceding discussion would
lead us to claim that

\begin{equation*}
L_n=\int d^3 x n_k [ F^{0k} (x) + e_0 x^k j^0 (x)]
\end{equation*}
is independent of time.  Since

\begin{equation*}
[j^0 (x), A^k (y)]_{x^0=y^0} = 0,
\end{equation*}
one would then conclude that

\begin{equation*}
i\, \langle 0 | [ L_n, A_k (y) ] |0\rangle = \frac {2}{3}\, n_k,
\end{equation*}
independent of $y^0$.  However, use of the spectral form for
$\langle 0|[A^u(x), A^v (y)|0\rangle$ quickly shows that the above relation cannot
be true except when $e_0=0$, and in fact the correctly computed right
hand side is $\frac{2}{3}\, n_k$ for $x^0=y^0$ so that $L_n$ generates
equal time constant gauge transformations, but for $x^0\neq y^0$ it
depends on all values of mass in the excitation spectrum of $A_k (x)$.
That there is no time independent generator of constant numerical
gauge transformations and hence no Goldstone theorem is a simple
restatement of the well established fact that there is no dynamical
reason why the physical photon should have zero mass\cite {12}.

Our argument has failed because in a theory which is not manifestly
covariant one is not able to demand that when

\begin{equation*}
\frac{\partial}{\partial x^u}\, \langle 0|[J^u (x), \Phi (y)]|0\rangle = 0
\end{equation*}
then

\begin{equation*}
\int d^3x\, \langle 0|[J^0 (x), \Phi(y)]|0\rangle
\end{equation*}
is independent of $(x^0 - y^0)$.  This is because causality
cannot be invoked to demonstrate that the other surface integrals that
arise in the application of Gauss's theorem to the above equation
vanish.  This statement should not be construed to mean that radiation
gauge electrodynamics is acausal.  It must be remembered that $A^u
(y)$ is an unphysical field so there is no reason to be concerned if
acausality appears in the study of some of its commutators.

Interacting Lorentz gauge electrodynamics has the field
equations $-\partial^2A^u = e\, j^u$ and $\partial_uj^u=0$ together with the
commutation relations given by equation \eqref{eq:4}.  It is easily found that

\begin{equation}
\label{eq:5}
i\, \langle 0|[A^u(x), A^v(y)]|0\rangle=(g^{uv}-\frac{\partial^u \partial^v}{\partial^2})[Z_3\Delta(x-y;0)+\int^\infty_{>0} d k^2 B(k^2) \Delta(x-y, k^2)+\frac{\partial^u \partial^v}{\partial^2} \Delta(x-y;0)
\end{equation}
Since we are dealing with a manifestly covariant theory the quantity

\[L_n=\int d^3x\, n_u[\partial_0A^u(x)-e_0x^u j^{0}(x)]\]
is independent of time as is indeed verified by direct computation which
yields the expected result

\begin{equation}
\label{eq:6}
i\, \langle 0|[L_n, A^\lambda(y)]|0\rangle = n^{\lambda}.
\end{equation}
Consequently there is a Goldstone theorem for interacting Lorentz
gauge electrodynamics.  There also is a strong point to be made from
this result. Note that \eqref{eq:6} is true even if $Z_3=0$ and there is no
physical zero mass particle.  If this should be the case the last term
of \eqref{eq:5} would be entirely responsible for the consistency of \eqref{eq:6}.
This term has occurred only because we have insisted on the somewhat
peculiar commutation relations \eqref{eq:4} and being purely gauge it does not
contribute to any physical amplitudes.  Thus as this simple example
illustrates, it is possible to have a Goldstone particle which is of
no interest whatsoever.  In electrodynamics we are fortunate because
the high degree of gauge invariance allows us to quantize in the
Coulomb gauge which has all unphysical modes removed.  In this gauge
it was easily found that no Goldstone theorem followed as an immediate
consequence of the structure of the field equations.  It is perhaps of
interest to note that after we are told there is a zero mass physical
photon as a direct result of solving the detailed dynamics for small
values of the coupling constant we can, of course, construct a
Goldstone theorem by applying the spatially integrated ``in'' or ``out''
field operator to the states in the usual way.  This is an example of
the inverse Goldstone theorem and is entirely after the fact.  Because
of these observations about the radiation gauge it is no surprise that
the zero mass Goldstone modes of Lorentz gauge electrodynamics are
purely unphysical.  Most theories of interest are not so highly gauge
invariant, and admit quantization only in a fully relativistic
manner.

Consequently it is necessary to examine the zero mass excitations very
carefully to ascertain whether they correspond to true particles which
appear in physical amplitudes.

So far we have used the method of broken symmetries in theories whose
Lagrangians are invariant under constant additions to some field.
Consequently we have been able to construct the ``broken symmetry''
states in a straight forward manner from the usual states and to
understand exactly what is meant by ``breaking'' a symmetry.  In short,
no results obtained are not more or less the direct consequences of
the normal field equations, and the methods we use while perhaps
amusing are an entirely unnecessary sophistication as far as solving
the problem at hand.  Indeed what we have done is something of a fraud
because the parameter of the ``broken symmetry'' appears in any formulae
in an entirely inert manner just because of the gauge invariance.  In
so far as any physical interpretation of results is concerned the
symmetry under consideration has not really broken at all.  Now we
wish to study theories which are less gauge invariant and for which
the parameter of the breaking appears in the Green's functions in a
physically significant manner.  Despite the non-trivial nature of
these more complicated theories we will be able to use the preceding
results as a guide to construct symmetry breaking theories where there
is no physical zero mass particle as a result of a Goldstone theorem.

Before proceeding to the more pertinent model it is beneficial as an
illustration of the above remarks to study the example given by
Goldstone \cite{1} with the Lagrangian

\[L=\varphi^u\partial_u\varphi+\frac{1}{2}\varphi^u\varphi_u+\frac{\mu^2_0}{2}\varphi^2-\frac{\lambda}{4}\varphi^2\varphi^2\;.\]
For the moment the number of components of the field $\varphi$ is left
unspecified.  We wish to solve this theory in the factorizable
approximation subject to the condition$\langle 0|\varphi(x)|0\rangle = n$.  It is
clear from the Lagrangian that the transformation $\varphi \rightarrow
\varphi+n$ is in no sense an invariance of the theory.  In order for
the broken symmetry condition to be consistent with the field
equations one finds the spectrum distorting condition $\mu^2_0=\lambda
n^2$.  Introducing the Green's function

\[g(x)=i\langle 0|(\varphi(x)
\varphi(0))|0\rangle -i\, n\, n\]
it is easily found that

\[g(p)=\frac{1}{p^2}[1-\frac{n n}{n^2}] + \frac{n\, n}{n^2}\, \frac{1}{p^2+2\lambda n^2}\;.\]
Note that in this theory the parameter $n$ appears in an entirely
non-trivial manner.  If $\varphi$ has only one component there is no
zero mass particle.  But if $\varphi$ has two (or more) components a
zero mass particle appears as the expression of the conserved current
$j^u=i \varphi^u q\, \varphi$ which yields the generator $Q=\int d^3 x
j^0$ and the Goldstone consistency condition $\langle 0|[Q,
\varphi(y)]|0\rangle = q\, n$.

Now we return to our problem, that of constructing a true broken
symmetry without a zero mass Goldstone boson.  The connection we have
established between gauge-invariance and the Goldstone theorem
suggests that we should try to find a failure of the gauge invariance
yields zero mass argument, and from this construct a broken symmetry
theory.  This is how the problem was initially solved, but we now take
the more direct route of starting with the Lagrangian

\begin{equation*}
L=-\frac{1}{2}F^{uv} (\partial_u A_v - \partial_v A_u)+\frac{1}{4} F^{uv} F_{uv} + \phi^u \partial_u \varphi + \frac{1}{2} \varphi^u \varphi_u + i e_0 (\varphi^u q \varphi) A_u.
\end{equation*}
Note that the current $j^u=i e_0 \varphi^u q \varphi$
satisfies the differential conservation law $\partial_u j^u=0$.  Any
approximation made on this theory will be required to respect this
conservation. We now impose the broken symmetry condition $i\, e\, q\,
\langle 0|\varphi|0\rangle = n \equiv \bigl(\begin{smallmatrix}n_1 \\n_2\end{smallmatrix}\bigr)$ and ask for the solution
of the above Lagrangian in the lowest factorizable approximation
\cite{13}.  In this approximation the complete self consistent Green's
function calculation is fully simulated by replacing the interaction
term by $(\varphi^u\, n)A_u$ and treating $\varphi$ as though it had
vanishing vacuum expectation.  We will thus for this presentation
avoid any complications by use of this device.  The resulting field
equations are

\begin{equation*}
F^{uv}=\partial^v A^v-\partial^v A^u
\end{equation*}

\begin{equation*}
\partial_v F^{uv}=\varphi^u n
\end{equation*}

\begin{equation*}
\varphi^u=-\partial^u\varphi-n A^u
\end{equation*}

\begin{equation*}
\partial_u\varphi^u=0\; .
\end{equation*}
Note that the current $\varphi^u n$ is conserved.  If $\varphi$ had
only one component these would essentially be the equation of one of
the models \cite {14} demonstrating that gauge invariance does not
always require zero mass.  Solving these equations in the radiation
gauge and taking (with no loss in generality) $n_2=0$ it follows that

\begin{equation*}
(-\partial^2+n^2_1)\varphi_1=0
\end{equation*}

\begin{equation*}
-\partial^2\varphi_2=0
\end{equation*}

\begin{equation*}
(-\partial^2+n^2_1) A^T_k=0
\end{equation*}

\begin{equation*}
\text{with}\; \nabla\cdot A^T_k=0 \; .
\end{equation*}

We thus see that the dimensional broken symmetry parameter $n$ plays
the role of a mass, serving to combine the two components of $A^T_k$
and the one component of $\varphi_1$, into the three components of a
massive vector meson.  The field $\varphi_2$ has been inert under this
process and while massless, is completely decoupled from the massive
excitations.  It is found rather directly that

\begin{equation}
\langle 0|[j^u(x), \varphi(y)]|0\rangle \equiv n_1\langle 0|[\varphi^u(x), \varphi(y)] 0\rangle
\end{equation}
so that

\begin{equation}
\langle 0|[j^u(x), \varphi_2]|0\rangle = 0
\end{equation}
and

\begin{equation}
\langle 0|[j^u(x), \varphi_1(y)]|0\rangle = n_1[-\partial^u+\frac{n^2[\lambda^u(\lambda\cdot\partial)+\partial^u]}{\partial^2+(\lambda\cdot\partial)^2}\Delta(x-y; n^2).
\end{equation}
Here we have introduced $\lambda^u=(1,0,0,0)$.

Note these results are consistent with differential current
conservation.  However, defining $Q(x^0)=\int d^3x\, j^0(x)$, we find
that

\[\langle 0|[Q(x^0), \varphi_1(y)]|0\rangle =-i n_1\cos(x^0-y^0)\,n_1 \;.\]
Consequently, because of the acausal nature of the
unphysical commutator there is no globally conserved charge and thus
there is no Goldstone theorem to invoke to argue that breaking the
symmetries requires that $\varphi_1$ excites massless particles.  It
should be appreciated that these results not only do not contradict,
but are actually required by the field equations which show that
$(\partial^2_0+n^2_1)Q=0$.  The fact that we are allowed the possibility
of quantizing this theory in the radiation gauge which is not
manifestly covariant allows us to strip the problem of unphysical
degrees of freedom and to see quickly to the heart of the matter.
However, since in most theories one is not allowed this possibility it
is good to point out what happens when these equations are examined in
the Lorentz gauge.  In this manifestly causal situation the global
charge $Q$ is of necessity independent of time.  We find that just as
in the case of interacting Lorentz gauge electrodynamics a pure gauge
zero mass part is added to the photon propagator and also becomes
associated with the operator $\varphi$.  It is this purely unphysical
mode which guarantees the consistency of the Goldstone theorem.

We have thus found a ``true'' broken symmetry theory which in its
relativistic form satisfies the Goldstone theorem through the
existence of unphysical modes, and in its equivalent radiation gauge
treatment avoids the restrictions of the Goldstone theorem because no
globally conserved symmetry operator exists.

The analysis of the behavior of this model in the radiation gauge
reveals why so many broken symmetry non-relativistic models fail to
have massless Goldstone bosons.  It is simply because when these
models have long range forces in their Hamiltonian (such as
$\frac{1}{r}$ Coulomb potentials) one must carefully check to see
whether the appropriate global generator is time independent.  In most
cases as an expression of the existence of these long range forces
some of the ``charge'' may oscillate in and out of the boundaries of any
box no matter how large and thereby negate any zero mass Goldstone
arguments.  The previously troublesome problem of the superconducting
electron gas at zero temperature with Coulomb interaction is explained
by this mechanism. \cite{7, 13}  In this case the plasma oscillations
have the property that their energy remains non-vanishing as
wavelength foes to infinity.  Consequently the Goldstone theorem is
not applicable to this problem.  For details see the paper by Lange.
There is a particularly nice point which our relativistic model
illustrates quite well.  In the superconducting model as soon as the
Coulomb interaction inserted in the Hamiltonian is screened ever so
slightly the massless Goldstone bosons appears.  This corresponds to
the fact that the surface integrals at infinity converge and
consequently that the global charge exists.  Note, that this
phenomenon would be particularly hard to explain if terms of the form
$n^u \delta^4 (k)$ carried the explanation of why zero mass particles
did not initially appear in this model.  Now the analogue of this
transition occurs in the model considered above when the smallest
amount of bare mass is associated with the field $A^u$.  In that case
a physical zero mass particle which serves to satisfy the Goldstone
criteria appears.

We conclude with a pessimistic note by pointing out that though the
above remarks lead to a much better understanding of the significance
of the Goldstone theorem, that it seems to me that we have a long way
to go before we find a relativistic model based on the method of
broken symmetries which has any real application to physical problems.
If we do succeed in finding such a model it is my feeling that it
probably will not have physical massless Goldstone bosons since the
``getting something for nothing'' aspect of the Goldstone theorem when
it applies to the physical mass spectrum very seriously limits the way
the broken symmetry can ``interact'' with the "normal" dynamics.

\section*{Acknowledgments}
I wish to thank Richard Hagen and Tom Kibble for the collaboration that made
the GHK paper possible, as well as for discussions over the years that have
taught me so much. This work is supported in part by funds provided by the US
Department of Energy (\textsf{DoE}) under \textsf{DE-FG02-91ER40688-TaskD}.

\end{document}